\documentclass[twocolumn,superscriptaddress,amssymb,amsmath,nobibnotes,aps,prx,longbibliography,showpacs]{revtex4}
\usepackage{graphicx}

\usepackage[colorlinks,citecolor=blue,linkcolor=Fuchsia]{hyperref}
\usepackage[usenames,dvipsnames,svgnames]{xcolor}

\usepackage{Jonasmacros}
\usepackage{bm}
\usepackage{mathptmx}
\usepackage{txfonts}
\newcommand{\tobedeleted}[1]{\textcolor{green}{#1}}
\renewcommand{\tobedeleted}[1]{\relax}

\begin{document}

\title{Detecting weak coupling in mesoscopic systems with a nonequilibrium Fano resonance}%

\author{S. Xiao}%
\affiliation{Department of Electrical Engineering, University at Buffalo, the State University of New York, Buffalo, NY 14260-1900, USA}
\author{Y. Yoon}%
\affiliation{Department of Electrical Engineering, University at Buffalo, the State University of New York, Buffalo, NY 14260-1900, USA}%
\author{Y.-H. Lee}%
\affiliation{Department of Electrical Engineering, University at Buffalo, the State University of New York, Buffalo, NY 14260-1900, USA}
\author{J. P. Bird}%
\email{jbird@buffalo.edu}
\affiliation{Department of Electrical Engineering, University at Buffalo, the State University of New York, Buffalo, NY 14260-1900, USA}
\affiliation{Graduate School of Advanced Integration Science, Chiba University, 1-33 Yayoi-cho, Inage-ku, Chiba 263-8522, Japan}
\author{Y. Ochiai}%
\affiliation{Graduate School of Advanced Integration Science, Chiba University, 1-33 Yayoi-cho, Inage-ku, Chiba 263-8522, Japan}
\author{N. Aoki}%
\affiliation{Graduate School of Advanced Integration Science, Chiba University, 1-33 Yayoi-cho, Inage-ku, Chiba 263-8522, Japan}
\author{J. L. Reno}%
\affiliation{CINT/Sandia National Laboratories, Dept. 1131, MS 1303, Albuquerque, NM 87185, USA}
\author{J. Fransson}%
\email{jonas.fransson@physics.uu.se}
\affiliation{Department of Physics and Astronomy, Uppsala University, Box 516, SE-751 21, Uppsala, Sweden}

\date{\today}

\begin{abstract}

A critical aspect of quantum mechanics is the nonlocal nature of the wavefunction, a characteristic that may yield unexpected coupling of nominally-isolated systems. The capacity to detect this coupling can be vital in many situations, especially those in which its strength is weak. In this work we address this problem in the context of mesoscopic physics, by implementing an electron-wave realization of a Fano interferometer using pairs of coupled quantum point contacts (QPCs). Within this scheme, the discrete level required for a Fano resonance is provided by pinching off one of the QPCs, thereby inducing the formation of a quasi-bound state at the center of its self-consistent potential barrier. Using this system, we demonstrate a form of \textit{nonequilibrium} Fano resonance (NEFR), in which nonlinear electrical biasing of the interferometer gives rise to pronounced distortions of its Fano resonance. Our experimental results are captured well by a quantitative theoretical model, which considers a system in which a standard two-path Fano interferometer is coupled to an additional, \textit{intruder}, continuum. According to this theory, the observed distortions in the Fano resonance arise \textit{only} in the presence of coupling to the intruder, indicating that the NEFR provides a sensitive means to infer the presence of weak coupling between mesoscopic systems.

\end{abstract}


\maketitle

\section{Introduction}
A central concept at the heart of physics is that extended systems may demonstrate rich behavior, not associated with their individual components but which arises when they are coupled to one another. Just a few different examples of this concept are provided by the natural bandstructures of periodic crystals, and their engineered counterparts in semiconductor superlattices \cite{Faist} and metamaterials \cite{Engheta}. In the emerging field of quantum information, the coupling of one system to another brings both benefits and disadvantages; on the one hand enabling sophisticated computations \cite{Nielsen}, while on the other giving rise to undesirable decoherence \cite{Joos}. Regardless of the ultimate application, in many cases there is a critical need to detect the coupling of different systems, especially under conditions where this coupling is weak. The objective of this work is to demonstrate the possibility of achieving such detection by exploiting the strong spectral sensitivity of Fano resonances \cite{Fano1961,Fano1986,Miroshnichenko,Lukyanchuk,Katsumoto} (FRs). Ubiquitous to both classical- \cite{Lukyanchuk} and quantum- \cite{Miroshnichenko,Katsumoto} wave systems, FRs are being explored for application in areas as diverse as nanoelectronics \cite{Katsumoto,Gores,Kobayashi,Song}, plasmonics \cite{Lukyanchuk,Hao,Verellen,Pakizeh,Liu}, metamaterials \cite{Plum,Papasimakis,Wu}, energy harvesting \cite{Fang}, and optics and nanophotonics \cite{Chao}. Here we explore their importance to the discussion of transport in quantum point contacts, in which we demonstrate a \textit{nonequilibrium} form of FR that provides an all-electrical scheme for the detection of weak quantum coupling in these mesoscopic devices.

\subsection{Fano resonances \& their extension to the nonequilibrium Fano resonance}
\begin{figure}
\begin{center}
\includegraphics[width=\columnwidth]{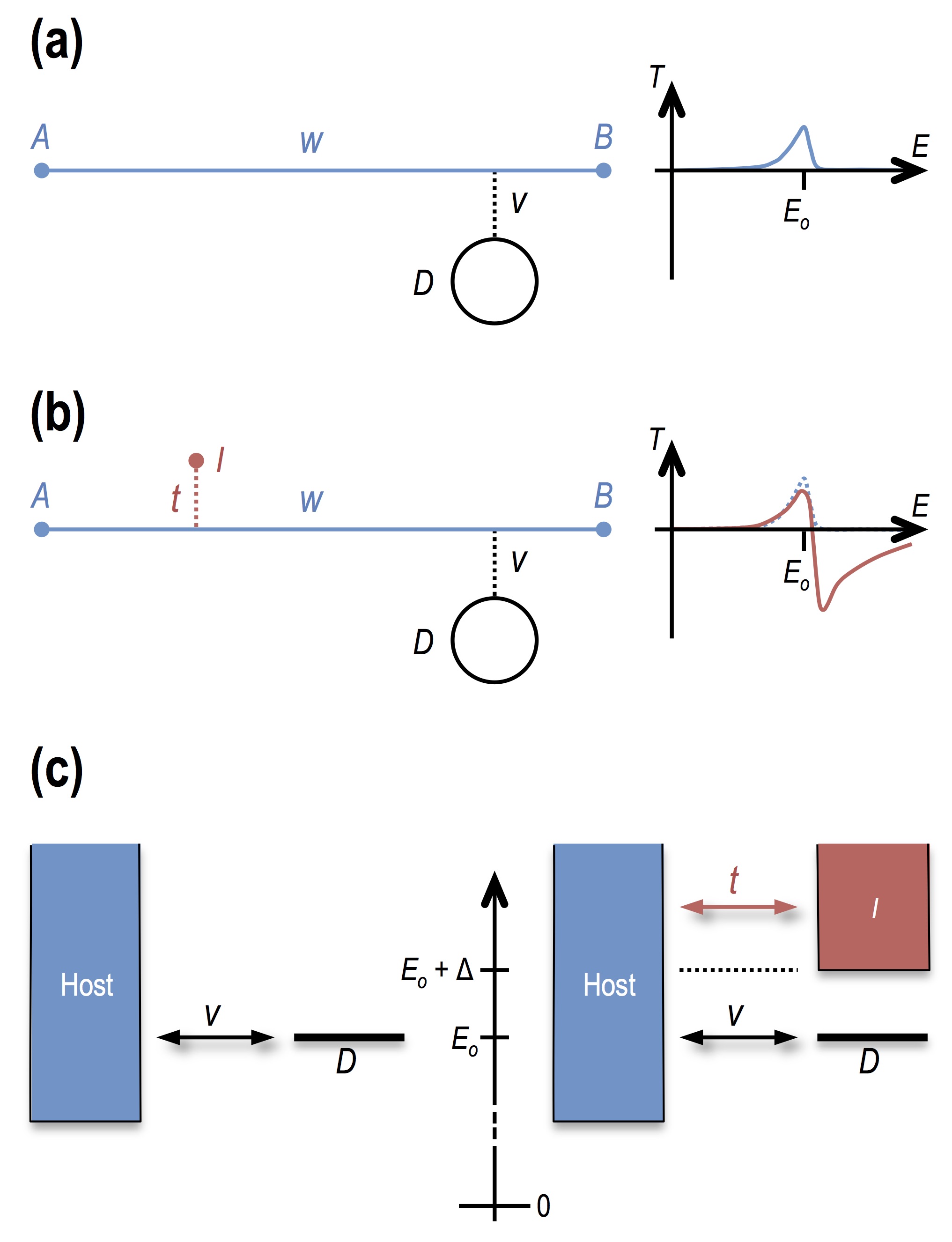}
\end{center}
\caption{(a) Shown left is a schematic representation of the standard two-path FR, while on the right we represent the possible variation of transmission as a function of energy ($E$) for such a system. The lineshape of the resonance reflects the relative values of the matrix elements $w$ and $v$, which in this case are shown to yield a weakly asymmetric peak. (b) The schematic on the left indicates the situation in which the Fano system of (a) is coupled to an additional, intruder, continuum ($I$). On the right we sketch the expected variation of $T(E)$ for this system. The dotted line represents the FR in the absence of the intruder, while the solid line indicates the lineshape distortion that can be induced by the coupling to $I$. (c) The left panel is a schematic representation of the energy alignment in the standard FR. The host continuum connects points $A$ and $B$ and is coupled to the discrete level ($D$) by the matrix element $v$. A FR results when mono-energetic waves are injected into the system with an energy close to $E_o$. The right panel shows the corresponding alignment relevant to discussions of the NEFR. Here, the host is coupled to the discrete level and the intruder ($I$), with respective matrix elements $v$ and $t$. The NEFR occurs when waves with a spread of energies, ranging from $E_o$ to $E_o+\Delta$, are simultaneously injected into the system to access both $D$ and $I$. The dotted line in the figure indicates the energy alignment of the intruder.}
\label{Fig1}
\end{figure}

FRs are observed in wave systems in which the transmission from an initial to a final state is governed by the interference between a continuum and a narrow level. Broadly realized in a variety of systems \cite{Fano1961,Fano1986,Miroshnichenko,Lukyanchuk,Katsumoto,Gores,Kobayashi,Song,Hao,Verellen,Pakizeh,Liu,Plum,Papasimakis,WuShvets,Fang,Chao}, the essential features of the Fano geometry are indicated schematically in Fig. \ref{Fig1}(a). This shows a problem in which waves propagate between points $A$ and $B$ (in some configuration space), with the transmission either occurring directly (matrix element $w$) or being mediated (with matrix element $v$) by a discrete level ($D$) that serves as an intermediate state. In this doubly-connected geometry, resultant wave interference causes the transmission ($T$) to exhibit a rapidly-varying resonant modulation (Fig. \ref{Fig1}(a), right panel), as the energy (or frequency) of the incoming wave is swept through that of the discrete level. The lineshape of the resonance takes a universal form whose profile is determined by an asymmetry parameter ($q$), which in turn is governed by the matrix elements $w$ and $v$ \cite{Fano1961,Fano1986}. Dependent upon the value of $q$ a variety of different lineshapes may be obtained, ranging from Lorentzian ($q = \infty$) and near-symmetric ($q\gg1$) forms, to fully antisymmetric ($q \sim 1$) and ``window'' resonances (or antiresonances, with $q = 0$). The capacity to manipulate the form of the FR via its asymmetry parameter, combined with the ability to very effectively modulate the transmission of an incident wave, are the features that make this phenomenon of such interest for use in the various applications alluded to above.
 
The manifestation of FRs in physical systems may be strongly enhanced under nonequilibrium conditions, with good examples being provided by the phenomenon of Raman scattering in doped semiconductors \cite{Cardona} and in carbon nanostructures \cite{Jorio}. In these materials, the interference scheme of Fig.\ref{Fig1}(a) is realized when a photon flux establishes a two-path transition from an initial state to a continuum: the first path involves the direct transmission between these states, while the second is mediated by a one-phonon Raman emission. In another example, a ``nonlinear Fano effect'' has been demonstrated in studies of the near-infrared absorption of self-assembled quantum dots \cite{Kroner}.  In these experiments, strong optical illumination was used to couple discrete states in the quantum dots to a two-dimensional continuum,
resulting in pronounced distortions of their photoabsorption peaks from a simple Lorentzian form. These resonances were instead well described by a model of Fano interference, in which the $q$-parameter could be parametrically varied by means of the incident laser power.

In the two examples discussed above, nonlinear optical excitation is used to \textit{establish} the two-path geometry required for Fano interference. In this work, however, we describe a different form of nonequilibrium Fano resonance (NEFR), which we observe in a mesoscopic system in which the two-path Fano interferometer is \textit{already} established under equilibrium conditions. By monitoring the distortions of its FR that arise when the system is subjected to strong nonequilibrium electrical driving (see Fig.\ref{Fig1}(b), right panel), we are able to infer the presence of coupling between the Fano interferometer and an \textit{additional} continuum. The realization of this phenomenon in an all-electrically-controlled scenario provides a useful contrast to prior demonstrations of optically-driven nonequilibrium Fano phenomena \cite{Cardona,Jorio,Kroner}, and confirms the capacity \cite{Tewari} to coherently manipulate carriers in mesoscopic devices under strongly nonlinear conditions. 

The generalized concept of the NEFR is indicated schematically in Fig. \ref{Fig1}(b). In this relatively-straightforward extension of Fig. \ref{Fig1}(a), the usual two-path Fano interferometer (see, also, the left schematic of Fig. \ref{Fig1}(c)) is modified by coupling it (with matrix element $t$) to an additional continuum ($I$). As we indicate in the right schematic of Fig. \ref{Fig1}(c), this \textit{intruder} \cite{Fransson} continuum is taken to be separated energetically from the discrete level by an energy detuning $\Delta$. In a situation in which mono-energetic waves are injected into the system to realize an equilibrium FR (at $E=E_o$), the intruder will therefore be energetically inaccessible and will consequently not participate in the resonant interference. However, if waves are injected into the same system with a spread of energy, chosen such that it matches the value of the detuning ($\Delta$), transmission via both the discrete level and the intruder can be activated simultaneously (see Fig. \ref{Fig1}(c)). Under such conditions, a three-path interferometer is established and it is this modification to the Fano geometry that results in the distortion of the resonance that we indicate schematically in Fig. \ref{Fig1}(b). The NEFR therefore provides a means to detect the coupling of the host to the intruder; put more simply, it allows us to detect the presence of \textit{hidden} components within a Fano system, even when they are undetectable in near-equilibrium transmission.

\subsection{Experimental implementation of the NEFR}
\begin{figure*}
\begin{center}
\includegraphics[width=\textwidth]{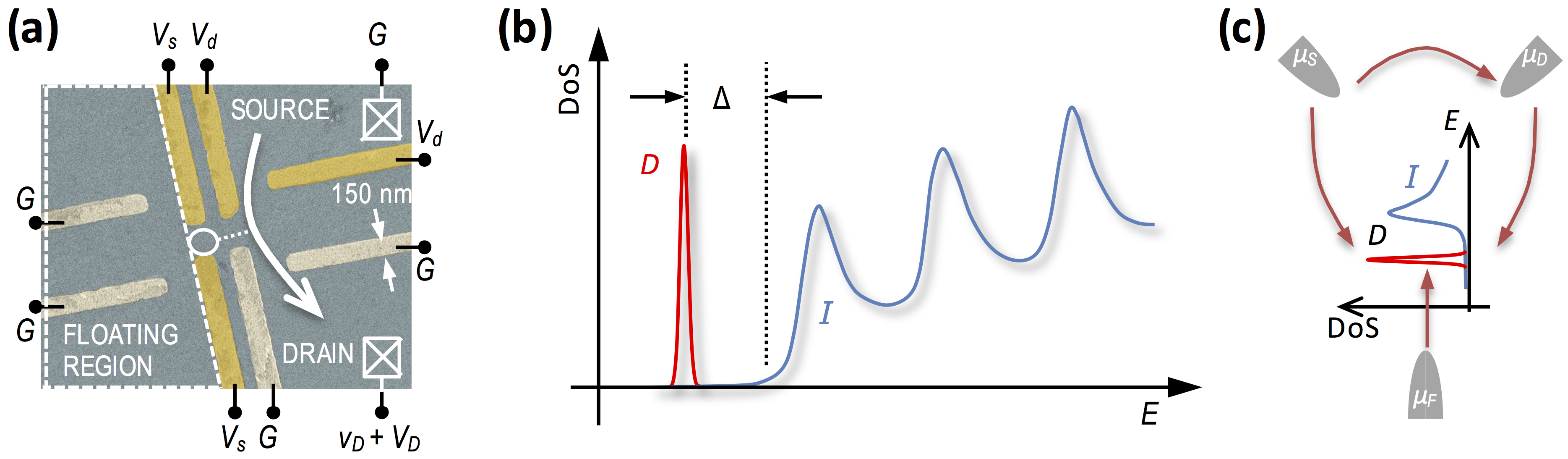}
\end{center}
\caption{(a) False color micrograph of a coupled-QPC device, in which the lighter-colored gates are held at ground potential ($G$) while the gold-colored gates are used to form the coupled QPCs. $V_s$ and $V_d$ represent the (DC) voltages applied the gates of the swept- and detector-QPCs, respectively. The white circle represents the natural quantum dot formed in the swept-QPC at pinch off, which is tunnel coupled (white dotted line) to electrons injected (solid white line with arrow) from the source of the detector. In measurements of the NEFR, the detector conductance is measured by superimposing a non-zero DC voltage ($V_D$) on top of a smaller AC component ($v_D$). (b) A schematic illustration of the density of states (DoS) in a QPC. This structure is only expected to be valid near pinch-off, where a spontaneously-formed discrete level ($D$) is present inside the QPC. The 1D subbands of the QPC represent the intruder and are separated from $D$ by an energy detuning $\Delta$. (c) Concept of the QPC-based intruder scheme. The Fano interferometer is created by coupling the DoS in (b) to the different reservoirs of the device (the source, drain and floating regions, at electrochemical potentials $\mu_S$,  $\mu_D$, and  $\mu_F$, respectively).}
\label{Fig2}
\end{figure*}

In our specific implementation of the NEFR, we implement the three-path interferometer of Fig. \ref{Fig1}(b) by exploiting the unusual properties of mesoscopic quantum point contacts \cite{Ferry} (QPCs) near pinch-off. QPCs are tunable electron waveguides that are typically realized by electrostatic gating \cite{Ferry} of a high-mobility two-dimensional electron gas (2DEG). In this approach, split-metal gates, separated by a nanoscale gap, are formed on top of the 2DEG substrate. By applying a suitable voltage to the gates the electrons directly underneath them can be depleted, leaving a narrow conducting channel within their gap. With the gate voltage adjusted such that the QPC is close to pinch-off, the charge within this channel can be reduced to the level of just a few electrons. In this ultra-low density limit, it has been argued theoretically that strong electron interactions can modify the self-consistent potential of the QPC, causing a natural quantum dot to spontaneously develop at its center \cite{Hirose,Rejec,Guclu,SongAhn,Welander}. While the existence of this self-consistent feature remains subject to debate (see the discussion at the end of this paper), a number of experiments nonetheless suggest that this scenario is correct \cite{Cronenwett,Ren,Klochan,Wu,Iqbal,Bird,Yoon2007,Yoon2009,Yoon2012,Fransson}. Among these include our own work \cite{Bird,Yoon2007,Yoon2009,Yoon2012,Fransson}, in which evidence for localized-state formation in pinched-off QPCs was provided by using this state to generate a FR in the (linear) conductance of a nearby (``detector-'') QPC. The measurement scheme used in these experiments is indicated in Fig. \ref{Fig2}(a), which shows two QPCs, separated by a few hundred nanometers, which are nonlocally coupled to one another through an intervening region of 2DEG. This latter region serves as a continuum of states and mediates coupling between the QPCs by means of wavefunction overlap \cite{Yoon2007,Yoon2012,Puller}. To induce the FR in the detector, the gate voltage ($V_s$) applied to the other (``swept-'') QPC is adjusted to align the discrete level of its quantum dot with the Fermi level in the common continuum. Under such conditions, a measurement of the linear conductance of the detector (i.e. a measurement performed by applying a vanishingly-small voltage across this QPC) exhibits a FR that results \cite{Fransson,Puller} from the interference of electron waves that are injected from the detector to the drain (solid line with arrow in Fig. \ref{Fig2}(a)), with those that reach the drain after first tunneling to and from the discrete level (dotted line in Fig. \ref{Fig2}(a)).

To introduce the role of the intruder into the above scenario for an equilibrium FR, it is necessary to consider somewhat more carefully the form of the density of states in the swept-QPC. Prior to pinch-off, where the QPC potential is well understood to be described by a parabolic saddle-like form \cite{Ferry}, the density of states consists of a set of (equally-spaced) quasi-continua. These correspond to the different one-dimensional (1D) subbands that mediate transport when the QPC is open, and which are responsible for the observation of its quantized conductance \cite{Ferry}. At pinch-off, however, the structure of the density of states should be markedly different, as we indicate in Fig. \ref{Fig2}(b). Due to the formation of a localized state, the lowest feature in the spectrum should correspond to a narrow peak. At the same time, the quasi-continua associated with the 1D subbands should be pushed above the Fermi level. Given this description, the connection to the intruder scheme of Fig. \ref{Fig1}(b) should be immediately apparent; the localized state formed within the QPC may serve as the discrete level ($D$) of a Fano scheme while the 1D subbands, separated from the discrete level by an energy detuning $\Delta$, play the role of the intruder. With the swept-QPC configured near pinch-off, and under near-equilibrium conditions, its localized state will lie near the Fermi level while the quasi-1D intruder will be energetically inaccessible at low temperatures. As we demonstrate here, however, a NEFR may be realized in this system by applying a suitable (nonlinear) source bias across the detector-QPC. Rather than probing the properties of the conductance near the Fermi level, as is done in small-signal transport studies, this allows us to simultaneously gain access both $D$ and $I$ (see Fig. \ref{Fig2}(c)) as required for the NEFR.

\section {Experimental methods}
\begin{figure}
\begin{center}
\includegraphics[width=\columnwidth]{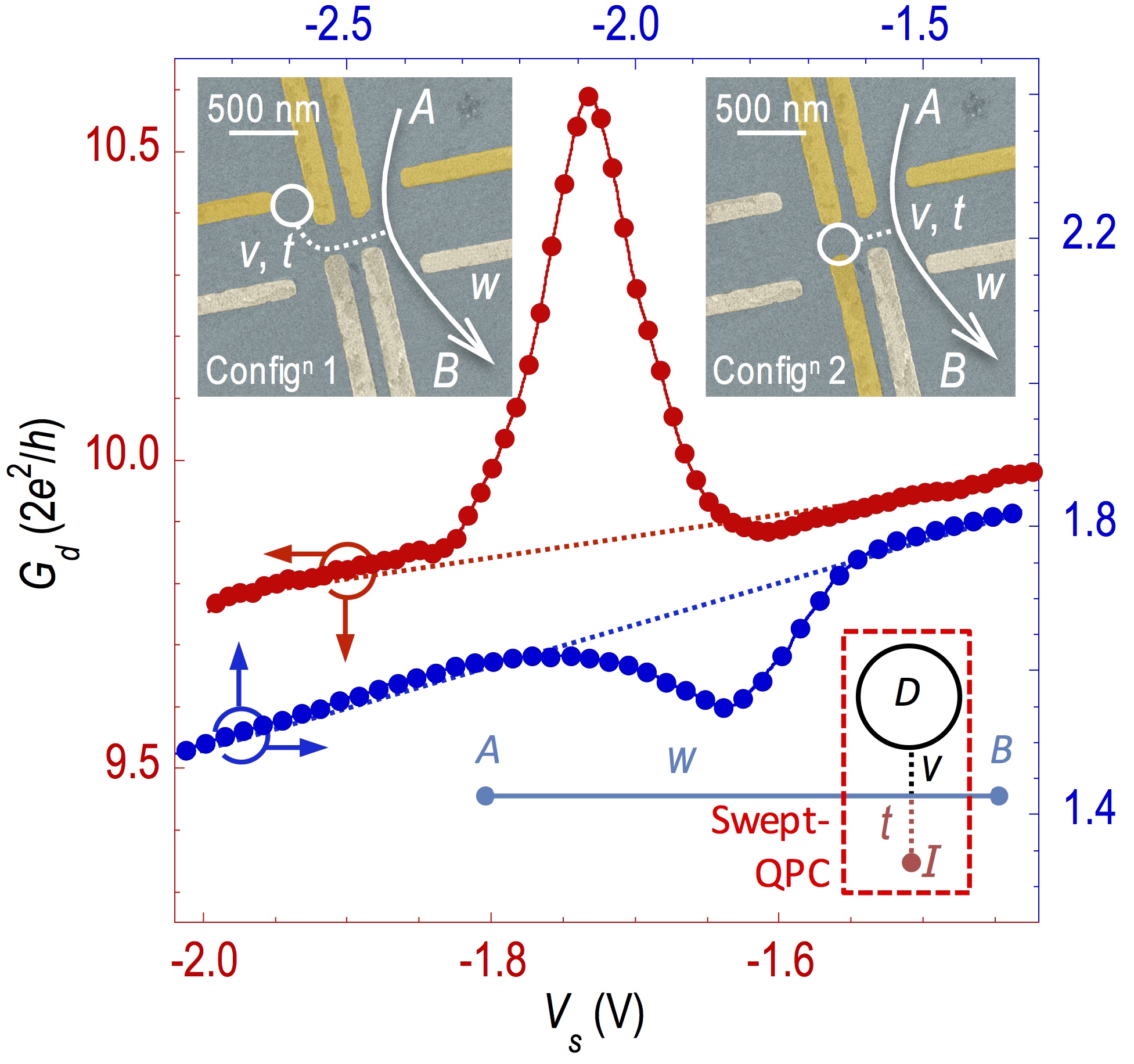}
\end{center}
\caption{The main panel shows measurements of the linear conductance ($V_D = 0$) of the detector QPC for the two coupled-QPC geometries identified as Configurations 1 $\&$ 2 in the false-color electron micrographs that form the upper insets to the figure. Red data correspond to the geometry identified as Configuration 1, while blue data correspond to Configuration 2. Dotted lines through the data represent the background subtracted from the raw conductance to obtain the resonant component.  The schematic at the bottom of the panel represents the realization of the intruder scenario in the coupled QPCs. The intruder and the discrete level are formed within the same QPC, as indicated by the red dotted line enclosing these two entities.}
\label{Fig3}
\end{figure}

Coupled QPCs were realized \cite{Yoon2007,Yoon2009,Yoon2012,Fransson} by electrostatic gating of high-quality GaAs/AlGaAs heterostructures (Sandia samples EA750 and VA0284, referred to hereafter as Devices 1 and 2). A 2DEG was formed in a 30-nm wide quantum well in these wafers, with a carrier density of $\sim 2\times10^{11}$ cm$^{-2}$ and mobility of $\sim3 \times 10^{6}$ cm$^{2}$/Vs. All experiments were performed at 4.2 K, a sufficiently low temperature to ensure coherent overlap between the coupled QPCs. AC conductance of the detector-QPC was measured with an RMS bias $v_D$ \textless 100  $\mu$V. The multi-gate geometry used to demonstrate the NEFR is indicated in Fig. \ref{Fig2}(a). By biasing specific pairs of gates, and leaving others grounded, coupled QPCs could be implemented in different configurations (Fig. \ref{Fig3}(a)). The lineshape of the FR exhibited in the linear conductance of the detector is known to be strongly configuration dependent \cite{Yoon2009}, indicating that varying the spatial arrangement of the two QPCs allows us to systematically manipulate the coupling elements ($w$ and $v$) appearing in the Fano problem. In previous work, we have largely focused on studies in which a small AC bias ($v_D$ in Fig. \ref{Fig2}(a)) was applied across the detector to determine its conductance near equilibrium. Here, however, we superimpose a larger DC voltage ($V_D$) upon $v_D$, thereby defining a non-zero energy window for transport (that gives rise to the NEFR when it contains both the discrete level and the intruder). While the voltages were actually applied to ohmic contacts at the edge of a 2DEG mesa (not indicated in Fig. \ref{Fig1}(b)), the small resistance ($\sim$20 $\Omega$) of these ungated regions ensured that the voltages were largely dropped across the detector QPC. Consequently, heating of the ungated 2DEG \cite{Kumar,Zakka} was not expected to be significant.

Important for the discussion that follows will be an understanding of the key energy scales associated with our system. Using the 2DEG density quoted above, we determine a Fermi energy of $\sim$6 meV in the ungated regions of the device. Separate bias-spectroscopy studies, on the other hand, indicate the energy spacing of the different 1D subbands that comprise the intruder to be in the range of 1 - 3 meV at pinch-off \cite{SongJungwoo}. Finally, an important parameter in Fig. 2(b) is the energy detuning ($\Delta$) between the discrete level and the edge of the intruder band. Self-consistent calculations based on spin-density functional theory \cite{Hirose} suggest that this energy, also, should be in the range of a few meV. We return to address this last point further below, in the light of our experimental results.

\section{Results}

\subsection{Experimental observations}
In Fig. \ref{Fig3} we demonstrate the form of the detector resonance obtained under conditions of linear transport (i.e. $V_D=0$) for two different coupled-QPC geometries that we refer to hereafter as Configurations 1 $\&$ 2 (see the upper insets to the figure). In both experiments, the variation of the detector conductance ($G_d$) is measured as the gate voltage ($V_s$) applied to the swept-QPC is used to pinch-off that structure. The resonant feature present in both curves is the FR of interest here and is superimposed upon a background (dotted lines in the main panel) that represents the direction electrostatic action of the swept-QPC gates on the detector \cite{Yoon2009}. In all subsequent analysis, this background is removed from the raw data, leaving only the resonant feature ($\Delta G_d$) in the detector conductance.

Turning to the issue of the lineshape of the resonances in Fig. \ref{Fig3}, these exhibit a pronounced influence of the specific coupled-QPC geometry. Specifically, in Configuration 1, the two QPCs are relatively far apart from one another and the FR is only weakly asymmetric. In Configuration 2, in contrast, the two QPCs are much closer to one another and form a stub-tuner like geometry that generates an antiresonance in the detector. This strongly configuration-dependent character of the detector resonance is consistent with the results of our previous, near-equilibrium, investigations \cite{Yoon2009}. The essential point is that, by varying the separation between the swept- and detector-QPC, we are essentially controlling the the relative magnitude of the matrix elements $w$ and $v$, with direct consequences for the Fano asymmetry parameter \cite{Yoon2009} Indeed, it is also worth noting that the antiresonance exhibited for Configuration 2 in Fig. \ref{Fig3} is reminiscent of that observed in studies in which such features have been induced in the conductance of narrow wires, by side-coupling them to intentionally-formed quantum dots \cite{Kobayashi2004}.

\begin{figure*}
\begin{center}
\includegraphics[width=\textwidth]{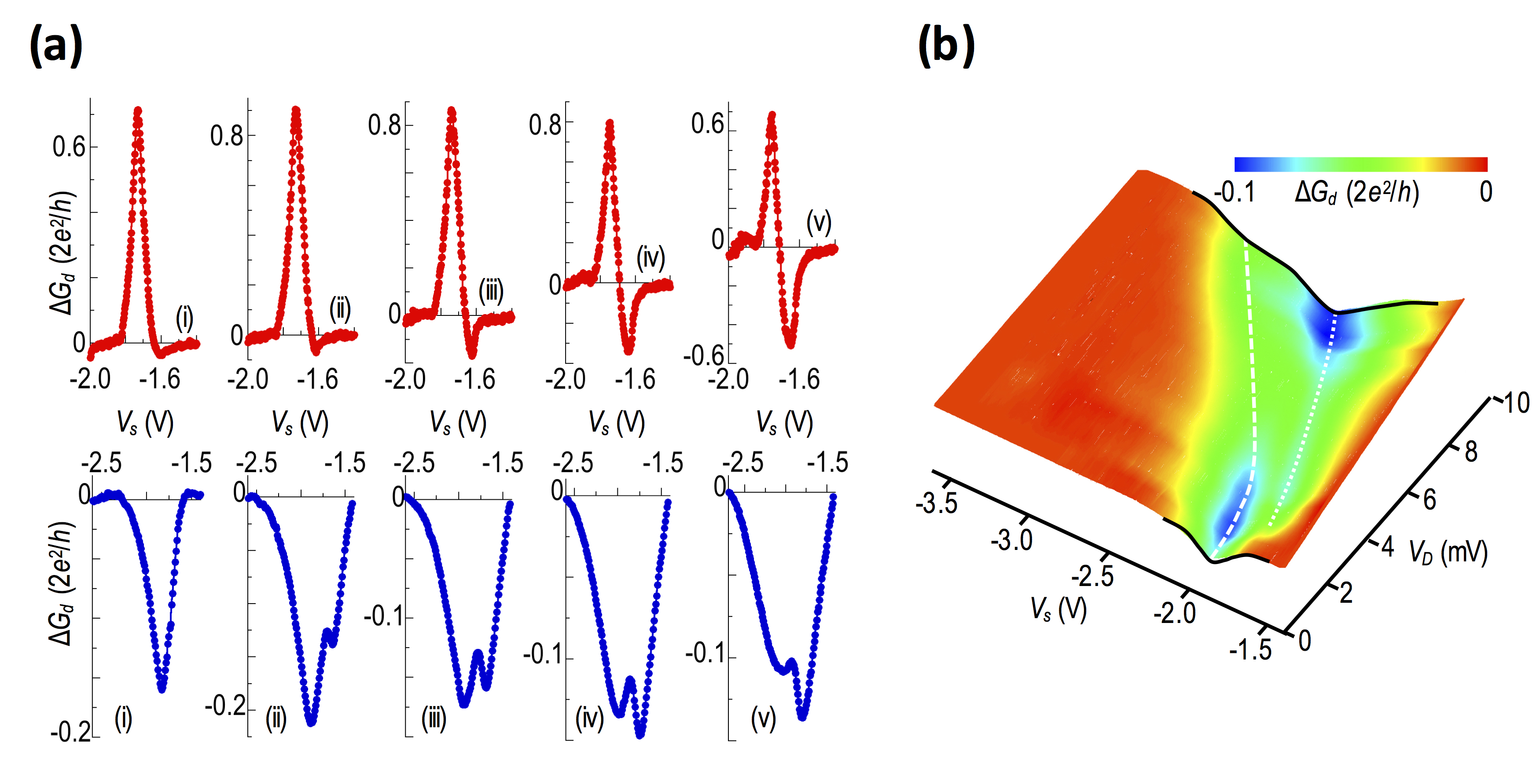}
\end{center}
\caption{(a) Measurements of the NEFR in the two different configurations. The red curves were measured in Device 1 in Configuration 1, while the blue curves were obtained in Configuration 2 for Device 2. (i) Red data: $V_D$ = 0 mV. Blue data: $V_D$ = 0 mV. (ii) Red data: $V_D$ = 3 mV. Blue data: $V_D$ = 2 mV. (iii) Red data: $V_D$ = 4 mV. Blue data: $V_D$ = 3 mV. (iv) Red data: $V_D$ = 5 mV. Blue data: $V_D$ = 4 mV. (v) Red data: $V_D$ = 6 mV. Blue data: $V_D$ = 5 mV. (b) Contour plot revealing the full variation of detector conductance as a function of $V_s$ and $V_D$, for Device 2 in Configuration 2. A monotonic background has been subtracted from $G_d$ to construct the contour. The white dashed line shows the evolution of the original antiresonance, present at $V_D = 0$, while the white dotted line represents the signature of the intruder that emerges for nonzero $V_D$.}
\label{Fig4}
\end{figure*}

In Fig. \ref{Fig4} we present our observations of the NEFR in Configurations 1 and 2, illustrating how the detector resonance is affected as $V_D$ is increased from zero. In both configurations we see that (Fig. \ref{Fig4}(a)), regardless of its initial form, the detector resonance develops a sharp dip on its less-negative gate-voltage side, the relative amplitude of which grows more pronounced as $V_D$ is increased. Recognizing the capacity of $V_s$ to act as a ``plunger'', that may sweep the local density of states of the QPC past the Fermi level, the presence of the dip at less-negative gate voltage than the main resonance indicates that this dip should be associated with a structure at higher energy than the discrete level. In fact, we will see shortly below that this structure is due to the edge of the 1D subbands indicated in Fig. \ref{Fig2}(b).

Figure \ref{Fig4}(b) shows the evolution of the detector resonance more systematically for Configuration 2, as a function of the two control voltages ($V_s$ and $V_D$). There are several noteworthy features of this contour, the first of which is the emergence of the additional dip (identified already in \ref{Fig4}(a)) on the ``high-energy'' flank of the original FR. This feature grows so strong by the maximal bias of 10 mV that it almost obscures the original resonance completely. Secondly, it is clear that, with increase of the bias $V_D$, both of these features (as indicated by the dotted and dashed lines) shift steadily to more-negative gate voltage. We have established previously \cite{Yoon2007,Yoon2009} that the FR exhibited in the linear conductance of the detector (at $V_D=0$) occurs soon after the swept-QPC pinches-off. The systematic shift of the white dashed line in Fig. \ref{Fig4}(b) therefore reflects the fact that, with larger source bias applied across a QPC, stronger (i.e. more-negative) gate biasing is needed to pinch it off \cite{Micolich}. The white dotted line in Fig. \ref{Fig4}(b) denotes the corresponding shift of the additional dip that appears on the high-energy side of the original FR and it is clear that this shows a similar dispersion to this original resonance. This is quite consistent with the picture of the density of states presented in Figure \ref{Fig2}(b) and allows us to interpret the separation between the two dips with the energy detuning ($\Delta$) between the discrete level and the 1D subbands. In a simple (non-self-consistent) picture of a ``rigid'' QPC potential, we might expect the $V_s$ separation of the two dips to remain constant as $V_D$ is increased. That this is not in fact the case indicates that the form of the self-consistent potential near pinch-off is modified by the detector bias. A description of this problem is beyond the scope of the current work. Nonetheless, it must be emphasized that the behavior shown in Figs. \ref{Fig3}(b) \& \ref{Fig3}(c) was not limited to these illustrative examples, but was reproduced in measurements performed on equivalent combinations of coupled QPCs, in both devices. It was also unaffected by varying detector conductance over a wide range ($1 \times 2e^2/h \leq G_d \leq 11 \times 2e^2/h$), confirming the idea that the resonance is driven by processes occurring within the swept-QPC.

A couple of further aspects of Fig. \ref{Fig4}(b) are worthy of clarification. Firstly, we note that the dispersion of the main resonance and its high-energy dip should not be confused with some kind of avoided crossing. Rather, as we have noted already, the separation between these features is reflective of the detuning ($\Delta$) between the discrete level and the 1D subbands. Even at $V_D=0$, this separation is not expected to vanish but rather to remain non-zero. Secondly, and more importantly, in the paragraph above we have emphasized the capacity of the nonlinear bias to influence the QPC potential. In our experiment, however, the bias in question ($V_D$) is applied to the \textit{detector}-QPC while the \textit{swept}-QPC is actually the one responsible for the observed resonance. In order to explain this apparent contradiction, it is necessary to consider the role of the ``floating region'' indicated in Fig. \ref{Fig2}(a). While this reservoir cannot draw any \textit{net} current from the supply, carriers may be injected into it from the detector as they exit it ballistically \cite{Folk,Ramamoorthy}. This process results in the appearance of a potential difference between the floating reservoir and the drain, which increases to reach a value sufficient to ensure that \textit{no} net charge is injected into the floating region. That is, application of voltage $V_D$ to the detector will result in the appearance of a potential drop across the swept-QPC, and it is this latter voltage that is responsible for the dispersion of the two resonances in Fig. \ref{Fig4}(b).

\begin{figure}
\begin{center}
\includegraphics[width=\columnwidth]{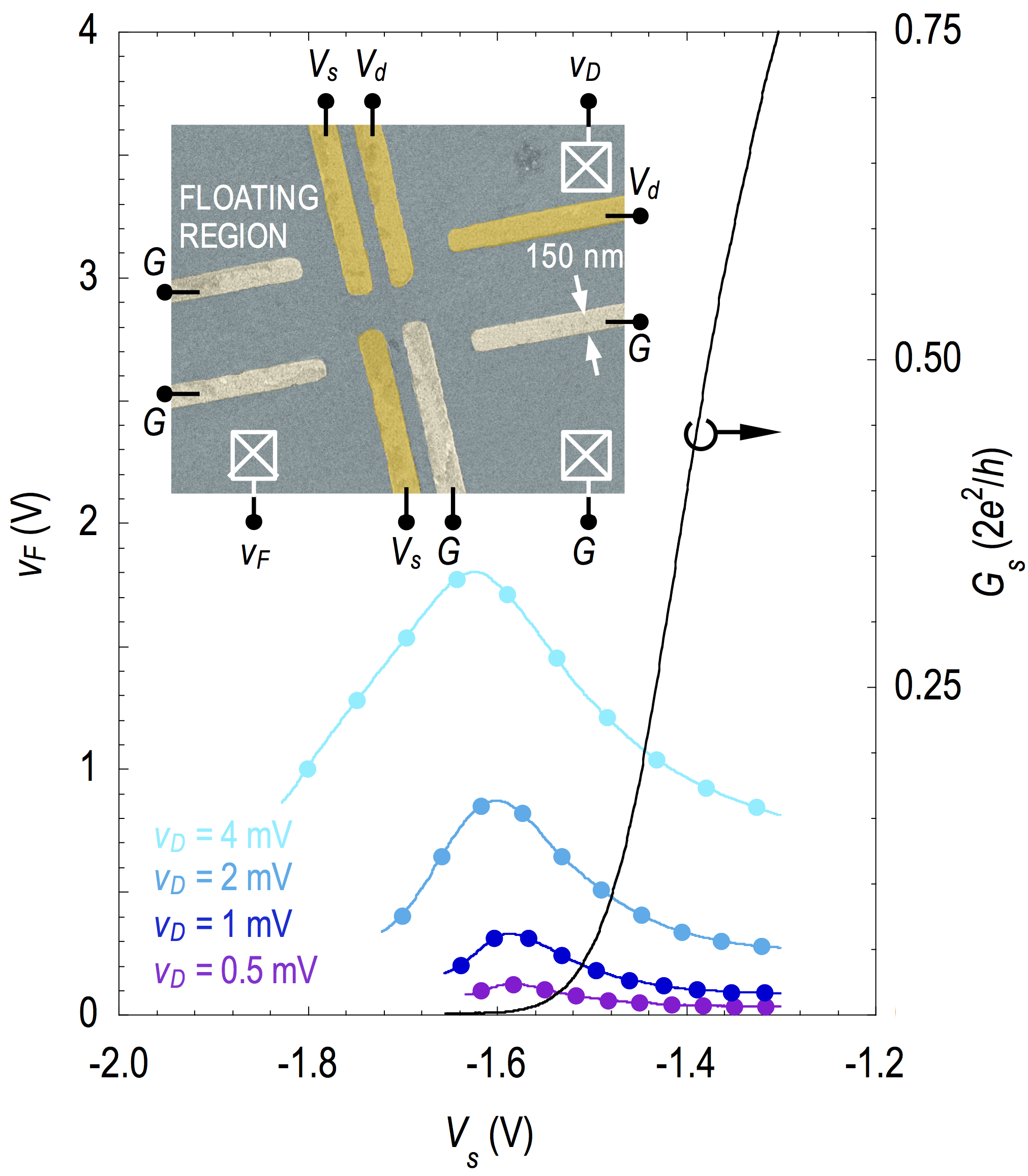}
\end{center}
\caption{Measurements of the floating-reservoir voltage ($v_F$) as a function of $V_s$, in the configuration indicated in the inset to the figure. The different curves were obtained by applying various AC voltages ($v_D$, RMS values indicated in the figure) across the detector, and then measuring the voltage of the floating reservoir relative to ground. Also shown is the corresponding variation of $G_s(V_s)$. }
\label{Fig5}
\end{figure}

In order to investigate the nature of the voltage that develops across the floating reservoir, we have performed a separate experiment using the configuration shown in Fig. \ref{Fig5}. As indicated in the inset to the figure, in this experiment we apply an AC voltage ($v_D$) of varying amplitude (RMS values are indicated in the figure) across the detector, and measure the resulting AC voltage ($v_F$) that develops at the floating electrode as $V_s$ is varied. Also shown for comparison in the figure is the dependence of the swept-QPC conductance ($G_s$) on $V_s$. Comparing the variation of $v_F$ with that of $G_s$, it is clear that a significant voltage develops at the floating reservoir as the swept-QPC approaches pinch-off. The maximum value of this voltage increases with increasing $v_D$, but it should be noted that it always remains less than the value of the supply voltage. As the swept-QPC pinches off, the reservoir voltage also decreases (although we are unable to observe its complete quenching due to the limited input impedance of our lock-in amplifier). Near pinch-off we therefore see that biasing of the detector can lead to the appearance of a significant voltage across the swept-QPC.

\begin{figure*}
\begin{center}
\includegraphics[width=\textwidth]{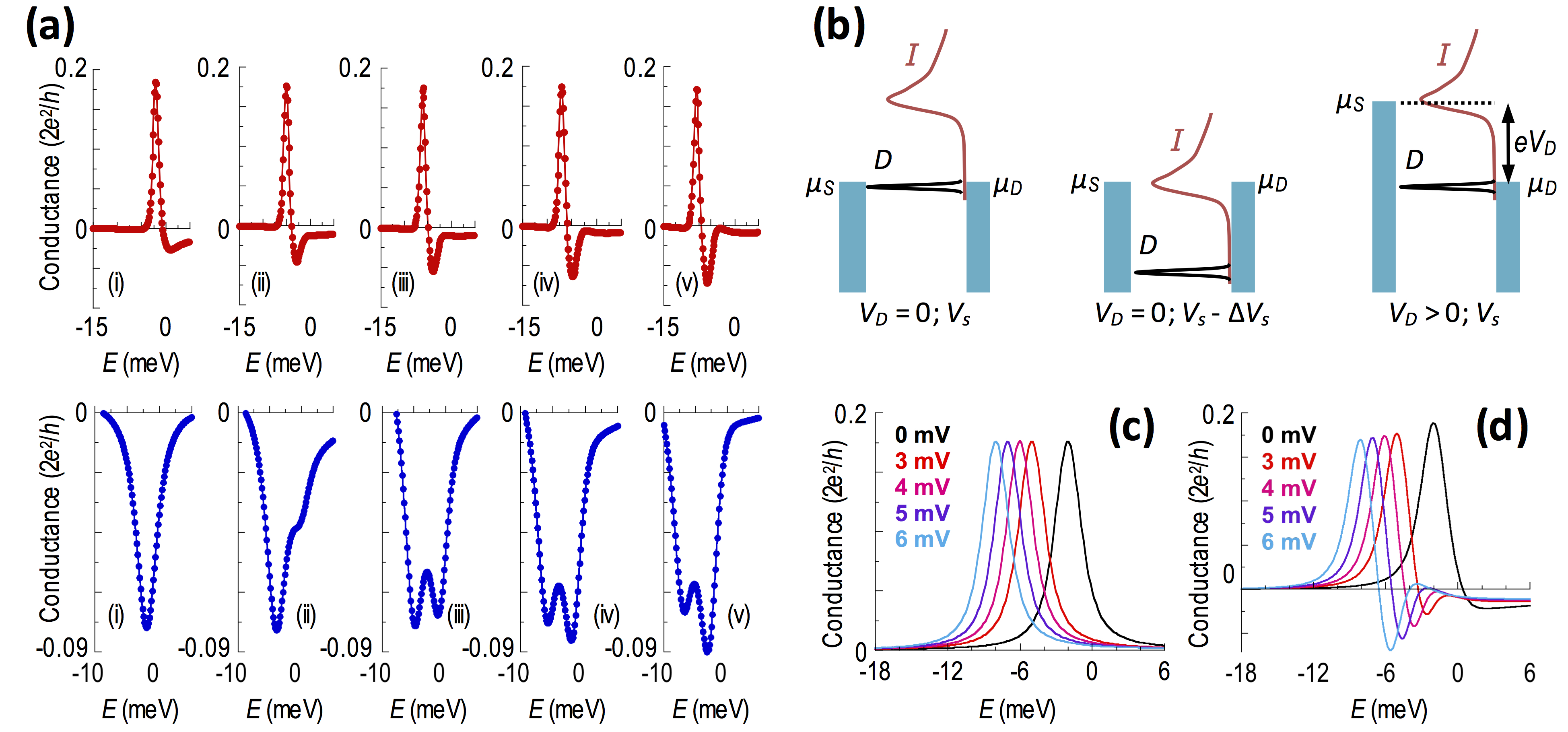}
\end{center}
\caption{(a) Model calculations of the NEFR for different system parameters. The top panels assume $t$ = 30- and $v$ = 30-meV, and are intended to reproduce the top set of panels of Fig. \ref{Fig4}(a). The lower panels, on the other hand, assume $t$ = $v$ = 40 meV, and capture the behavior for the bottom set of panels of Fig. \ref{Fig4}(a). (b) Schematic illustrations showing the level alignments in the system under different conditions. The left and center panels are for thermal equilibrium ($v_D$ = $V_D$ = 0), where either the localized state ($D$, left) or the edge of the 1D-subbands (center) is aligned with the reservoir chemical potentials. Shown right, however, is the nonequilibrium situation, where the applied voltage ($V_D$) opens up an effective energy window that may be used to simultaneously couple to both $D$ and $I$. (c) NEFR computed for $t$ = 0, corresponding to the usual two-path FR shown in the upper schematic of Fig. \ref{Fig1}(a). The calculations assume $v$ = 30 meV and are performed for different $V_D$ (values indicated). (d) Similar NEFR ($v$ = 30 meV, values of $V_D$ again indicated), but for $t$ = 30 meV.}
\label{Fig6}
\end{figure*}

\subsection{Theoretical modeling of experiment}
The detailed variations of the detector resonance in Fig. \ref{Fig3} are reproduced well by a theoretical model that we have developed, and which attributes them to a form of NEFR. While the detailed derivation of this model is provided in the appendix to this paper, the essential idea of our approach is to model the system as a localized state and a 1D band, embedded in a junction formed by three separate regions of 2DEG. These are distinguished by their electrochemical potentials $\mu_S$,  $\mu_D$, and  $\mu_F$, in the source, drain, and floating reservoirs, respectively (refer to Figs. \ref{Fig2}(a) \& \ref{Fig2}(c)). The Hamiltonian for this system is solved by introducing appropriate matrix elements to describe the coupling between the different components of the system. In this way we are able to compute the dependence of the detector conductance on both energy and applied source bias ($V_D$), and to therefore model the behavior in experiment.

In Fig. \ref{Fig6}(a), we present the results of calculations of the detector conductance as a function of energy (equivalent to variation of $V_s$) and $V_D$. The resulting figures clearly capture the essential features of our experiment (see the corresponding panels of Fig. \ref{Fig3}(b)), reproducing the sharp dip that appears on the lower-energy side of the resonance when $V_D$ is applied. The amplitude of the resonances obtained from the model is roughly half that of those observed in experiment. Given the relative simplicity of our model, which does not attempt to treat the specific microscopic details of our experimental system, we consider this degree of agreement to be satisfactory. The correspondence between experiment and theory allows us to attribute our observations to a NEFR, involving a mechanism that is indicated in Fig. \ref{Fig6}(b). Here we indicate the influence of using a variation of the swept-QPC gate voltage ($V_s$) to scan the local density of states of this QPC past $\mu_S$ and $\mu_D$ (compare the left and center panels). Close to thermal equilibrium ($v_D$ = $V_D$ = 0), and at the low temperatures that we consider here, this alignment may be achieved separately for either the localized state (Fig. \ref{Fig6}(b), left panel) or the 1D continuum (center panel), but not simultaneously for both features. This situation is overcome at non-zero $V_D$, which allows $\mu_S$ and $\mu_D$ to be separately aligned with $I$ and $D$, for an appropriate bias that matches $\Delta/e$ (Fig. \ref{Fig6}(b), right panel). In other words, the nonequilibrium conditions allow us to \textit{complete} the three-path Fano interferometer, revealing the coupling to the intruder. This point is made clear by comparing the influence of nonequilibrium biasing on the usual two-path (Fig. \ref{Fig6}(c)), and three-path (Fig. \ref{Fig6}(d)), FR. Figure \ref{Fig4}(c) was obtained for an intruder coupling $t$ = 0 (i.e. no intruder influence) and shows only a rigid shift in the position of the detector resonance as $V_D$ is varied, without the appearance of any dip. As we have described already, a similar shift is also seen in experiment (see Fig. \ref{Fig4}(b), for example), in which the position of the detector resonance shifts to more-negative $V_s$ as $V_D$ is similarly increased. The shift is apparent again in the results of Fig. \ref{Fig6}(d), obtained this time with nonzero coupling to the intruder, although the most dramatic feature here is a strong distortion of the FR due to the presence of the intruder. The distortion appears on the high-energy side of the resonance, in agreement with the results of experiment and consistent with the idea that the source of the intruder is indeed the 1D subbands of the pinched-off (swept-) QPC.

While the good agreement between experiment and theory in Figs. \ref{Fig4} \& \ref{Fig6} provides strong support for our interpretation of the NEFR, this agreement is dependent upon the choice of model parameters (most notably $t$ and $v$) whose values cannot be determined from first principles. For this reason, it is important to provide some kind of justification for the values used for these parameters in the various curves in Fig. \ref{Fig6}. The essential point here is that, in order to fit the results obtained in Configuration 1, we require smaller parameter values ($t$ = $v$ = 30 meV) than those needed to fit the data for Configuration 2. This appears to at least be reasonable, since in Configuration 1 the two QPCs are farther apart than in Configuration 2 and so the matrix elements $t$ and $v$ should be smaller in this case. The other point that must be emphasized is that the various nonlinear resonances in Figs. \ref{Fig4} \& \ref{Fig6} \textit{cannot} be fitted using the usual Fano asymetry ($q$-) parameter. This is clearly obvious for the data obtained in Configuration 2, in which the NEFR exhibits a ``double-dip'' structure that is completely inconsistent with any known Fano form \cite{Fano1961}. Even the nonlinear data obtained in Configuration 1, which appear reminiscent of the classic $q\sim1$ lineshape, however, do not conform to the universal Fano form. This point was emphasized previously in our earlier study of the "magnetically-tuned'' FR \cite{Fransson}, where we showed that the dip that develops due to the intruder cannot be fitted by the same $q$-value needed to describe the peak due to the discrete level. In the context of the nonlinear experiment of interest here, this point can be understood by appealing to the results of our theoretical model. Most notably, in Eqs. (\ref{eq-NEFR1}) and (\ref{eq-NEFR2}) of the appendix we present expressions for the separate contributions to the NEFR from the discrete level and the intruder, respectively. While the former contains a term that resembles the usual Fano lineshape, the latter modifies this lineshape under nonequilibrium conditions, so that the overall resonance can no longer be expected to be described by Fano's universal form.

\section{discussion}

\subsection{Connection to earlier work on Fano-interference schemes}
Recently, there has been much attention devoted to the observation of a \emph{nonlinear FR}, in studies of the near-infrared photoabsorption in self-assembled quantum dots \cite{Kroner}. In this experiment, strong mono-energetic laser excitation was used to reveal clear evidence of Fano interference in the quantum-dot absorption process. The interference arose from the presence of two distinct pathways for excitonic transitions, the first involving direct electron-hole excitation within the dot, while the second involved a process in which this transition was mediated by an intervening continuum. Physically, the source of the continuum was a wetting layer in close proximity to the quantum dot, and the matrix element for transmission through it could be increased by increasing the laser power. In other words, the role of the nonlinear excitation in this experiment was to form the usual two-path Fano geometry. More recently, a similarly-tunable nonlinear FR has been considered \cite{Zhang} for hybrid nanostructures, comprised of semiconductor quantum dots coupled to metallic nanoparticles. In such systems, the excitonic modes of the quantum dots and the plasmonic ones of the nanoparticles correspond, respectively, to the discrete and continuum states of a Fano scheme. These components are then again coupled to one another through nonlinear excitation to give rise to the FR. Both of these examples \cite{Kroner,Zhang} are therefore different to the NEFR discussed here, in which the Fano system is already formed under equilibrium and the nonlinear electrical excitation (with non-monoenergetic waves) is instead used to reveal its coupling to an additional system (i.e. the intruder).

Our demonstration of the NEFR represents another example of a double-continuum FR, with strong conceptual overlap with recent work on plasmonic Fano systems \cite{Arju}. There, the possibility of ``continuum-state competition'' has been discussed, in which one continuum can significantly influence the Fano interference exhibited by another. This possibility was actually established in our earlier experimental study \cite{Fransson}, in which we demonstrated the capacity of our coupled-QPC scheme to provide a realization of the intruder. To observe the influence of this feature in near-equilibrium transport, it was necessary in that work to apply a strong magnetic field perpendicular to the plane of the device. By causing wavefunction compression of the different QPC states, this allowed us to reduce the detuning between the discrete level and the intruder, thereby allowing the signature of the latter to emerge in the detector conductance. The approach here, in contrast, is very different, making use of nonequilibrium biasing to reveal the coupling to the intruder, without the need for a magnetic field. This capacity to implement the NEFR by all-electrical means moreover provides us with a useful scheme to perform spectroscopy of the intruder system (see below), something that was not directly possible in the magnetic-field studies.

Finally we note that, in recent theoretical work \cite{Shapiro}, a generalized description of FRs in the presence of dissipation was developed. The essential conclusion reached by the authors of that work was that the dissipation results in a modified FR, whose lineshape now features an additional Lorentzian contribution. The situation in our experiment is very different, since we consider how the FR lineshape is modified by the introduction of the intruder, when it provides an additional path for \textit{coherent} interference in the system. The resulting lineshape in this case reflects the specific form of the density of the states of the one-dimensional intruder, and we do not treat the role of dissipation at all in our theoretical model. The essential agreement that we achieve between experiment and our more-restricted theory suggests that, at least in the low-temperature regime that we consider, dissipation does not play a primary role in influencing the lineshape of the NEFR.

\subsection{Does a localized state really form in QPCs?}
While the emphasis in this study has been on the use of coupled QPCs to demonstrate the NEFR, our results also have additional impact in terms of their relevance to ongoing discussions, as to whether a localized state can in fact form spontaneously in a QPC at pinch-off. The most recent contribution to this debate has come in the form of measurements of the electronic magnetization of QPCs from nuclear magnetic resonance \cite{Kawamura}. In that work, the authors inferred a smooth change in magnetization as a function of QPC barrier height and used this result to conclude that no localized state is formed within the QPC. Our experiments clearly contradict this interpretation; most notably, the NEFR relies at its core on the existence of a discrete level in the pinched-off QPC. From our prior work \cite{Yoon2007,Yoon2009,Fransson}, which has shown that the quantitative features of the detector resonance are reproduced systematically in multiple QPCs, fabricated in different heterostructures, we can moreover rule out a ``chance'' impurity as the source of this localized state. We also emphasize an important difference between our nonlocal transport investigations and the local measurements of QPC conductance that are usually made in any experiment (including that of \cite{Kawamura}). Specifically, we have established previously \cite{Yoon2007,Yoon2009,Yoon2012,Fransson} that the FR exhibited by the detector is observed immediately \textit{after} the swept-QPC pinches-off. As such, our measurements of coupled systems provide us with a means to access information on the QPC electronic structure in a regime where the conductance has vanished and which is therefore inaccessible in local investigations of individual QPCs.

Finally, we point out that our measurements of the NEFR provide us with a technique to perform a spectroscopy of the local density of states of the pinched-off QPC. More specifically, our numerical simulations of the NEFR indicate (see Fig. \ref{Fig4}) that the nonlinear distortion of the FR should onset once the energy window opened by the nonlinear bias ($V_D$) becomes comparable to the detuning ($\Delta$, see Fig. \ref{Fig2}(b)) between the discrete level and the intruder. From the data of Fig. \ref{Fig3}, it would appear that the relevant value for this detuning is in the range of a few meV. As noted earlier, this estimate is consistent with the results of self-consistent calculations based on spin-density functional theory \cite{Hirose}.

\section{conclusions}
In conclusion, we have demonstrated an approach to detect weak (tunnel) coupling between quantum systems by making use of a nonequilibrium Fano resonance that differs from the usual implementations of this phenomenon. The essential idea of this approach is to exploit the strong sensitivity of the Fano resonance to the presence of additional transmission pathways, as a means to identify the coupling of some host to an intruder. Crucially, our nonequilibrium scheme allows us to detect the presence of ``hidden'' components within some system, even while they remain ``invisible'' in near-equilibrium transport. For a proof-of-concept demonstration of this phenomenon, we implemented an electron-wave interferometer from mesoscopic quantum point contacts. Against a backdrop of continued theoretical interest \cite{Huang,Shapiro} in the potential applications of FRs, our experiment serves to demonstrate the rich physical behavior that can be realized by extending the Fano interferometer beyond its usual two-path form.

\begin{acknowledgements}
The experimental research in the group of JPB at Buffalo was supported by the U.S. Department of Energy, Office of Basic Energy Sciences, Division of Materials Sciences and Engineering under Award DE-FG02-04ER46180. Epitaxial growth of the high-quality 2DEG layers was performed by JLR at the Center for Integrated Nanotechnologies, a U.S. Department of Energy, Office of Basic Energy Sciences user facility. Sandia National Laboratories is a multi-program laboratory managed and operated by Sandia Corporation, a wholly owned subsidiary of Lockheed Martin Corporation, for the U.S. Department of Energy's National Nuclear Security Administration under contract DE-AC04-94AL85000. JF acknowledges support from the Swedish Research Council. 
\end{acknowledgements}

\appendix*
\section{Modeling the nonequilibrium Fano resonance}

We model the system by considering a localized state (LS) and a 1D band, embedded in the junction formed by three separate regions of two-dimensional electron gas (2DEG). These three reservoirs are distinguished by their electrochemical potentials $\mu_S$, $\mu_D$, and $\mu_F$, corresponding to the source, drain, and floating reservoirs, respectively. Electrons in the source and drain reservoirs are connected directly to each other via the rate $w$. They are also coupled to the LS with rates $v_S$ and $v_D$, and to the 1D band via rates $t_S$ and $t_D$. The LS and 1D band are also coupled to the third reservoir (\emph{F}) with rates $v_F$ and $t_F$, respectively. The Hamiltonian for this setup is written as

\begin{align}
\Hamil=&
	\sum_{\bfk\in\chi}E_{\chi\bfk}n_{\chi\bfk}
	+\dote{0}n
	+\sum_\bfq\dote{\bfq}n_\bfq
\nonumber\\&
	+
	\left\{
		\sum_{\bfk\in L,\bfk'\in R}w\csdagger{\bfk}\cs{\bfk'}
		+
		\sum_{\bfk\in\chi}\csdagger{\bfk}
		\Bigl(
			v_\chi\dc{}
			+
			t_\chi a_\bfq
		\Bigr)
		+
		H.c.
	\right\}
	.
\end{align}
Here, $E_{\chi\bfk}$ is the energy of an electron in reservoir $\chi=S,D,F$, whereas $\dote{0}$ and $\dote{\bfq}$ denote the energies of electrons in the LS and the 1D band. The operators $\cs{\bfk}$, $d$, and $a_\bfq$ destroy electrons in the reservoirs, the LS, and the 1D band, respectively, and $n$ is the number operator. We omit any reference to spin in the present case. 

The expression for the stationary charge current flowing between the source and drain can be written
\begin{align}
I(V)=&
	\frac{ie}{\hbar}\sum_{\bfk\in S}
		\biggl(
			f_S(\dote{\bfk})Q^>(\dote{\bfk})
			+
			f_S(-\dote{\bfk})Q^<(\dote{\bfk})
		\biggr),
\end{align}
where the lesser/greater form, $Q^{</>}(\omega)$, of the propagator $Q(z)$ describes the physics of the electron bath in the drain and of the electrons in the swept-QPC, as well as the interactions in the model. $f_\chi(\omega)=f(\omega-\mu_\chi)$ is the Fermi function at the chemical potential $\mu_\chi$.

\subsection{Standard Fano resonance}
Ignoring, for now, the presence of the propagating 1D states in the swept-QPC, we can factorize $Q^{</>}(\omega)$ according to
\begin{align}
Q^{</>}=&
	\Bigl(
		v_S+wv_D\sum_\bfk g^r_\bfk
	\Bigr)
	\calG^r\Sigma^{</>}\calG^a
	\Bigl(
		v_S^*+w^*v_D^*\sum_{\bfk'}g^a_{\bfk'}
	\Bigr)
\nonumber\\&
	+
	\biggl[
		|w|^2
		+
		2\re
		\Bigl\{
			wv_D
			\Bigl[
				v_S^*+w^*v_D^*\sum_{\bfk'}g^a_{\bfk'}
			\Bigr]
			\calG^a
		\Bigr\}
	\biggr]
	\sum_\bfk g^{</>}_\bfk
\end{align}
We introduce the notation
\begin{subequations}
\begin{align}
A=&
	v_S+wv_D\sum_{\bfk\in D}\frac{1}{\omega-E_\bfk},
\\
B=&
	\pi wv_D\sum_{\bfk\in D}\delta(\omega-E_\bfk),
\end{align}
\end{subequations}
and define the Fano factor $q=A/B$. We further notice that
\begin{subequations}
\begin{align}
wv_D\sum_{\bfk\in D} g^{</>}_\bfk(\omega)=&
	(\pm i)2f_D(\pm\omega)B,
\\
|w|^2\sum_{\bfk\in D} g^{</>}_\bfk(\omega)=&
	(\pm i)2f_D(\pm\omega)\frac{|B|^2}{\Gamma^D/2}.
\end{align}
\end{subequations}
Finally, we also have $\calG^r(\omega)\Sigma^{</>}(\omega)\calG^a(\omega)=\Sigma^{</>}(\omega)|\calG^r(\omega)|^2$, where
\begin{subequations}
\begin{align}
\Sigma^{</>}(\omega)=&
	(\pm i)\sum_{\chi=S,D,F}f_\chi(\pm\omega)\Gamma^\chi,
\\
|\calG^r(\omega)|^2=&
	\frac{1}{|\omega-\dote{0}-\Sigma^r(\omega)|^2}
	=
	\frac{1}{(\Gamma/2)^2}\frac{1}{\dote{}^2+1}.
\end{align}
\end{subequations}
Here, we have defined $\dote{}=(\omega-\dote{0}-\re\Sigma^r)/(\Gamma/2)$, with $\Gamma=-2\im\Sigma^r(\omega)=\sum_\chi\Gamma^\chi$ and $\Gamma^\chi=2\pi\sum_{\bfk\in\chi}|v_\chi|^2\delta(\omega-\dote{\bfk})$. In this way we obtain
\begin{align}
\calG^r(\omega)\Sigma^{</>}(\omega)\calG^a(\omega)=&
\nonumber\\&\hspace{-3cm}=
	(\pm i)\frac{f_S(\pm\omega)\Gamma^S+f_D(\pm\omega)\Gamma^D+f_F(\pm\omega)\Gamma^F}{(\Gamma/2)^2}
	\frac{1}{\dote{}^2+1}
	.
\end{align}
Substituting these expressions into that for the current, we find that
\begin{align}
I(V)=&
	4\frac{e}{\hbar}|B|^2
	\sum_{\bfk\in S}
		\Bigl(f_S-f_D\Bigr)
		\biggl(
			\frac{\Gamma^D}{\Gamma^2}
			\frac{q^2+1}{\dote{}^2+1}
			+
			\frac{1}{\Gamma^D}
			+
			\frac{2}{\Gamma}
			\frac{\dote{}q-1}{\dote{}^2+1}
		\biggr)
\nonumber\\&
	+
	4\frac{e}{\hbar}|B|^2
	\sum_{\bfk\in S}
		\Bigl(f_S-f_F\Bigr)
			\frac{\Gamma^F}{\Gamma^2}
			\frac{q^2+1}{\dote{}^2+1}
		.
\label{eq-classicFano}
\end{align}
In the limit $\Gamma^S/\Gamma^D\ll1$, $\Gamma^F/\Gamma^D\ll1$ where $\Gamma\approx\Gamma^D$, the current reduces to
\begin{align}
I\approx
	\frac{4e}{\hbar}\sum_{\bfk\in S}
		\frac{|B|^2}{\Gamma}
		\Bigl(f_S(\dote{\bfk})-f_D(\dote{\bfk})\Bigr)
		\frac{(\dote{}+q)^2}{\dote{}^2+1},
\end{align}
which gives the typical Fano interference formula for a single LS in the propagating pathway. Using non-equilibrium Green functions we have, hence, established a straightforward route to obtain the classic expression \cite{Fano1961,Fano1986} for the Fano resonance.

\subsection{Nonequilibrium Fano resonance}
To describe the nonequilibrium Fano resonance we now include the propagating states in the swept-QPC and employ the same method as above to obtain
\begin{widetext}
\begin{align}
Q^{</>}=&
	\biggl[
		|w|^2
		+
		|t_S|^2|t_D|^2\Bigl|\sum_\bfq a^r_\bfq\Bigr|^2
	\biggr]
	\sum_{\bfk\in D} g^{</>}_\bfk
	+
	2\re\Biggl\{
		\biggl[
			v_S+wv_D\sum_{\bfk\in R} g^r_\bfk+wt_St_D^*\sum_{\bfq;\bfk\in D}a^r_\bfq g^r_\bfk
		\biggr]
		\calG^r
		\biggl[
			w^*v_D^*+w^*t_S^*t_D\sum_{\bfq'} a^a_{\bfq'}
		\biggr]
		\Biggr\}
		\sum_{\bfk'\in D} g^{</>}_{\bfk'}
\nonumber\\&
	+
	\biggl[
		v_S+wv_D\sum_{\bfk\in D} g^r_\bfk+wt_St_D^*\sum_{\bfq;\bfk\in D} a^r_\bfq g^r_\bfk
	\biggr]
	\calG^{</>}
	\biggl[
		v_S^*+w^*v_D^*\sum_{\bfk'\in D} g^a_{\bfk'}+w^*t_S^*t_D\sum_{\bfq';\bfk'\in D} g^a_{\bfk'}a^a_{\bfq'}
	\biggr]
\nonumber\\&
	+
	|t_S|^2
	\biggr[
		1
		+
		2|t_D|^2\re\sum_{\bfq;\bfk\in D}g^r_\bfk a^r_\bfq
	\biggr]
	\sum_{\bfq'} a^{</>}_{\bfq'}
	+
	2\re\Biggl\{
		\biggl[
			v_S+wv_D\sum_{\bfk\in D} g^r_\bfk+wt_St_D^*\sum_{\bfq;\bfk\in D}a^r_\bfq g^r_\bfk
		\biggr]
		\calG^r
		w^*t_S^*t_D\sum_{\bfk'\in D} g^r_{\bfk'}
		\Biggr\}
		\sum_{\bfq'} a^{</>}_{\bfq'}.
\label{eq-tildeQ}
\end{align}
\end{widetext}
Here we define the parameters
\begin{subequations}
\begin{align}
\tilde{A}=&
	v_S+wv_D\sum_{\bfk\in D}\re\, g_\bfk^r+wt_St_D^*\sum_{\bfq;\bfk\in D}\re\, a^r_\bfq g^r_\bfk,
\\
\tilde{B}=&
	-wv_D\sum_{\bfk\in D}\im\, g^r_\bfk-wt_St_D^*\sum_{\bfq;\bfk\in D}\im\, a^r_\bfq g^r_\bfk,
\end{align}
\end{subequations}
and $\tilde{q}=\tilde{A}/\tilde{B}$. Using $\calG^{</>}=\Sigma^{</>}|\calG^r|^2$ with $\calG^r$ and $\Sigma^{</>}$ defined as in the previous case, we can write the third contribution (second line) to $Q^{</>}$ as $4|\tilde{B}|^2(|\tilde{q}|^2+1)\Sigma^{</>}/[(\dote{}^2+1)\Gamma^2]$, which has the same functional appearance as the corresponding contribution to the standard Fano resonance. Similarly, we can write the first contribution (first line) to $Q^{</>}$ as
\begin{align}
(\pm i)&
	2f_D(\pm\omega)
	\frac{
		\Bigl|
			-wv_D\sum_\bfk\im\, g^r_\bfk-wt_St_D^*\sum_{\bfq\bfk}\im\, a^r_\bfq g^r_\bfk
		\Bigr|^2
	}{
		-\sum_\bfk|w|^2\im\, g^r_\bfk
	}
\nonumber\\=&
	(\pm i)4f_D(\pm\omega)
		\frac{|\tilde{B}|^2}{\Gamma^D},
\end{align}
under the condition that $\re\sum_\bfk g^r_\bfk\approx0$, which holds for the metallic state in the 2DEG. By the same token, we can then also write the second contribution (first line) as
\begin{align}
(\pm i)&
	4f_D(\pm\omega)\re\{\tilde{B}(\tilde{q}-i)\calG^r\tilde{B}^*\}=
	(\pm i)8f_D(\pm\omega)\frac{|\tilde{B}|^2}{\Gamma}
	\frac{\dote{}\tilde{q}-1}{\dote{}^2+1}.
\end{align}
Inserting the first three contributions from Eq. (\ref{eq-tildeQ}) into the current we obtain
\begin{align}
I_{LS}=&
	4\frac{e}{\hbar}
	|B|^2\sum_{\bfk\in S}\Bigl(f_S-f_D\Bigr)
	\biggl(
		\frac{\Gamma^D}{\Gamma^2}\frac{\tilde{q}^2+1}{\dote{}^2+1}
		+
		\frac{1}{\Gamma^D}
		+
		\frac{2}{\Gamma}
		\frac{\dote{}\tilde{q}-1}{\dote{}^2+1}
	\biggr)
\nonumber\\&
	+
	4\frac{e}{\hbar}|B|^2\sum_{\bfk\in S}\Bigl(f_S-f_F\Bigr)
	\frac{\Gamma^F}{\Gamma^2}\frac{\tilde{q}^2+1}{\dote{}^2+1}
	,
\label{eq-LSFano}
\end{align}
which is formally the same expression as that for the standard Fano resonance in Eq. (\ref{eq-classicFano}).

Finally, the fourth and fifth terms (third line) in Eq. (\ref{eq-tildeQ}) can be written as
\begin{align}
	\re&
	\Biggl\{
		|t_S|^2
		+
		2|t_S|^2|t_D|^2\sum_{\bfq;\bfk\in D}g^r_\bfk a^r_\bfq
\nonumber\\&
		+
		2B(\tilde{q}-i)
		\calG^r
		w^*t_S^*t_D\sum_{\bfk'\in D} g^r_{\bfk'}
	\Biggr\}
	\sum_{\bfq'} a^{</>}_{\bfq'}
	=
	\calA_S
	\sum_\bfq a^{</>}_\bfq
	.
\end{align}
Since the propagator $a_\bfq$ for the propagating states in the swept-QPC couples to electrons in both the source and drain reservoirs we have
\begin{subequations}
\begin{align}
a^{r/a}_\bfq(\omega)=&
	\frac{1}{\omega-\dote{\bfq}\pm i\gamma/2},
\\
a^{</>}_\bfq(\omega)=&
	(\pm i)\sum_\chi f_\chi(\pm\omega)\gamma_\chi
	|a^r_\bfq(\omega)|^2,
\end{align}
\end{subequations}
with $\gamma=\sum_\chi\gamma_\chi$ and $\gamma_\chi=2\pi\sum_{\bfk\in\chi}|t_\chi|^2\delta(\omega-E_\bfk)$. Summing over the momenta $\bfq$, we obtain (setting $\kappa_\pm^2=2N_0(\omega-\dote{\sw}\pm i\gamma/2)$)
\begin{align}
\sum_\bfq|a^r_\bfq(\omega)|^2=&
	\int_0^{\infty}\frac{1}{|\omega-\dote{q}+i\gamma/2|^2}\frac{dq}{2\pi}
	=
	\frac{N_0}{\gamma}
	\biggl(
		\frac{1}{\kappa_+}
		+
		\frac{1}{\kappa_-}
	\biggr).
\end{align}
The corresponding contribution to the current therefore becomes
\begin{align}
I_{1D}=&
	\frac{e}{\hbar}\calA_S\sum_{\bfq\bfk\in S}
	\biggl(
		\gamma_D\Bigl(f_S-f_D\Bigr)
		+
		\gamma_F\Bigl(f_S-f_F\Bigr)
	\biggr)
	|a_\bfq^r|^2.
\label{eq-1DFano}
\end{align}

\subsection{Differential conductance}
We assume that $\mu_S=\mu_D+eV$, $\mu_F=\mu_D+\alpha_FeV$, and $\mu_D=0$, with $0\leq\alpha_F\leq1$, so that the (differential) conductance from Eqs. (\ref{eq-LSFano}) and (\ref{eq-1DFano}) for can be written at low temperatures as
\begin{subequations}
\begin{align}
\frac{dI_{LS}}{dV}=&
	4\frac{e^2}{\hbar}|B|^2
	\sum_{\bfk\in S}
		\Biggl[
			\biggl(
				\frac{\Gamma^D}{\Gamma^2}\frac{\tilde{q}^2+1}{\dote{}^2+1}
\nonumber\\&
				+
				\frac{1}{\Gamma^D}
				+
				\frac{2}{\Gamma}\frac{\dote{}\tilde{q}-1}{\dote{}^2+1}
				+
				\frac{\Gamma^F}{\Gamma^2}\frac{\tilde{q}^2+1}{\dote{}^2+1}
			\biggr)
			\delta(\dote{\bfk}-eV)
\nonumber\\&
			-
			\alpha_F\delta(\dote{\bfk}-\alpha_FeV)
			\frac{\Gamma^F}{\Gamma^2}\frac{\tilde{q}^2+1}{\dote{}^2+1}
		\Biggr],
\\
\frac{dI_{1D}}{dV}=&
	\frac{e^2}{\hbar}
	\calA_S\sum_{\bfq\bfk\in S}
	\Biggl[
		\Bigl(
			\gamma_D+\gamma_F
		\Bigr)
		\delta(\dote{\bfk}-eV)
\nonumber\\&
		-
		\alpha_F
			\gamma_F\delta(\dote{\bfk}-\alpha_FeV)
	\Biggr]
	|a^r_\bfq(\dote{\bfk})|^2
	,
\end{align}
\end{subequations}
respectively. Physically, the parameter $\alpha_F$ accounts for the voltage drop between the floating and right reservoirs, and is dependent on charge accumulation around the swept-QPC.
In correspondence to to the experimental situation, we take the limit $\Gamma^S/\Gamma^D\ll1$, $\Gamma^F/\Gamma^D\ll1$ where $\Gamma\approx\Gamma^D$, and $\gamma_D/\gamma_F\ll1$. The conductances are then dominated by
\begin{subequations}
\begin{align}
\frac{dI_{LS}}{dV}=&
	4\frac{e^2}{\hbar}\sum_{\bfk\in S}
		\frac{|B|^2}{\Gamma}
		\frac{(\dote{}+\tilde{q})^2}{\dote{}^2+1}
		\delta(\dote{\bfk}-eV),
\label{eq-NEFR1}
\\
\frac{dI_{1D}}{dV}=&
	\frac{e^2}{\hbar}
	\calA_S\gamma_F
	\sum_{\bfq\bfk\in S}
	\biggl(
		\delta(\dote{\bfk}-eV)
\nonumber\\&
		-
		\alpha_F\delta(\dote{\bfk}-\alpha_FeV)
	\biggr)
	|a^r_\bfq(\dote{\bfk})|^2
\label{eq-NEFR2}	.
\end{align}
\end{subequations}
This shows that the contribution from the LS generates a standard Fano-resonance in the conductance when the $\dote{0}$ is swept through the chemical potential, behavior described by the ratio $(\dote{}+\tilde{q})^2/(\dote{}^2+1)$. The contribution from the intruder generates, on the other hand, a signature with the characteristic shape of the 1D density of electron states, as can be seen from the presence of the integrated 1D Green function $\sum_\bfq|a^r_\bfq(\omega)|^2\sim\sum_{s=\pm}1/\sqrt{\omega-\dote{\sw}+is\gamma/2}$. The difference in the parentheses signifies that the 1D density of electron states is available only when there is a finite voltage drop across the swept-QPC, that is when this QPC is in a non-equilibrium state.

\end{document}